# Large Magnetic-Field-Induced Strains in Sintered Chromium Tellurides

Yuki Kubota[1], Yoshihiko Okamoto[1,2,a], Tomoya Kanematsu[1], Takeshi Yajima[2], Daigorou Hirai[1], and Koshi Takenaka[1]

[1]*Department of Applied Physics, Nagoya University, Nagoya 464-8603, Japan*
[2]*Institute for Solid State Physics, University of Tokyo, Kashiwa 277-8581, Japan*
[a)] yokamoto@issp.u-tokyo.ac.jp

Sintered samples of $Cr_3Te_4$ and $Cr_2Te_3$ are found to show large strains accompanied by large volume changes under a magnetic field. In $Cr_3Te_4$, volume increases of $\Delta V/V$ = 500–1170 ppm by applying a magnetic field of 9 T are observed over the entire temperature range below 350 K. At room temperature, the $\Delta V/V$ value exceeds 1000 ppm, which is considerably larger than the maximum values reported for Cr-based magnets thus far and is comparable to the room-temperature value of forced-volume magnetostriction in invar alloys. $Cr_2Te_3$ show a large $\Delta V/V$ of 680 ppm when applying a magnetic field of 9 T at 200 K. Both samples display particularly large volume increases around the Curie temperature, where they also show negative thermal expansion due to microstructural effects, suggesting that the cooperation between anisotropic lattice deformation associated with the magnetic ordering and microstructural effects is essential for the manifestation of the large magnetic-field-induced volume changes.

The strain that magnetic materials experience when a magnetic field is applied has long been the subject of research. The magnetostriction of iron and nickel, which show linear strains of several tens of ppm, is the most classical example. Subsequent studies have led to the development of giant magnetostrictive materials, such as Terfenol-D, which can realize large magnetostrictions exceeding 2000 ppm.[1] Such materials have been put into practical use as ultrasonic transducers and actuators. The strains in the ferromagnetic metals mentioned above are caused by the alignment of ferromagnetic domains in magnetic fields and are therefore essentially directional with little volume changes and are saturated at high magnetic fields where the magnetization saturates. In contrast, invar alloys have displayed large volume changes induced by magnetic fields, known as forced volume magnetostriction.[2] The magnitude of such a change was reported to be $\Delta V/V$ = 70 ppm under a magnetic field of 0.5 T at room temperature, which is much larger than the forced volume magnetostrictions in iron and nickel.[3] In invar alloys, large spontaneous magnetostriction caused by the strong magneto-volume effect cancels the thermal expansion due to lattice vibrations, resulting in low thermal expansion.[2] This strong magneto-volume effect also plays a major role in the emergence of large volume changes under magnetic fields.

In recent years, magnetic-field-induced strains accompanied by large volume changes have been discovered in several Cr-based magnets, which have been infrequently thought of as candidate magnetostrictive materials. Large volume increases reaching $\Delta V/V$ = 700 ppm under a magnetic field of 9 T were observed in $LiInCr_4S_8$ just below the Néel temperature of $T_N$ = 24 K and in $AgCrS_2$ at $T_N$ = 42 K.[4,5] These volume increases are smaller than those in invar alloys, but remain reasonably large. $ZnCr_2Se_4$ and $Cr_2Te_3$ have also been reported to exhibit large linear strains in magnetic fields, although it is unclear whether they are accompanied by volume changes.[6,7] Most of these Cr-based magnets are antiferromagnetic insulators, unlike invar alloys,[8,9] suggesting that the large magnetic-field-induced volume changes in Cr-based magnets are caused by a new mechanism based on the characteristic features of Cr spins, such as magnetic ordering accompanied by structural distortion, strong spin-lattice coupling, and geometrical frustration. Therefore, further studies on the magnetostructural properties of Cr-based magnets are expected to lead to the emergence of novel correlated phenomena between their magnetism and volume. In addition, in Cr-based magnets, the large volume changes are realized in a narrow temperature window far below room temperature, because the magnetic phase transition plays an essential role. Therefore, if we use them as magnetostrictive materials, they are only available for limited applications.

In this letter, we focus on the magnetic-field-induced strain in Cr-Te binary compounds, namely, $Cr_3Te_4$ and $Cr_2Te_3$. As shown in Fig. 1(a), both compounds crystallize in the NiAs-based crystal structures, where the metal atoms form a primitive hexagonal structure with stacked triangular lattices. In $Cr_3Te_4$ and $Cr_2Te_3$, completely occupied Cr layers and Cr deficient layers with Cr vacancies stack alternately. In $Cr_3Te_4$, the Cr atoms are linearly aligned in a Cr-deficient layer, resulting in monoclinic $C2/m$ symmetry.[10,11] In contrast, $Cr_2Te_3$ has a honeycomb-type order made of Cr atoms and vacancies in a Cr-deficient layer, resulting in trigonal $P\bar{3}1c$ symmetry.[12] $Cr_3Te_4$ and $Cr_2Te_3$ were reported to be metallic ferromagnets with Curie temperatures of $T_C = 320$ and 180 K, respectively.[11-14]

Recently, Cr-Te binary compounds, including $Cr_3Te_4$ and $Cr_2Te_3$, have been intensively studied from the perspective of room-temperature ferromagnets with tunable $T_C$ values.[15-17] However, they are not simple ferromagnets. The magnetization of $Cr_3Te_4$ slightly decreases at $T_t = 80$ K, far below $T_C$, when the temperature is decreased in a low magnetic field. This decrease was proposed to be due to the change in spin structure from ferromagnetic to canted antiferromagnetic.[12,18,19] In $Cr_2Te_3$, neutron scattering data suggests that the Cr spin moment in the Cr deficient layers is negligibly small and opposite to the direction of the net magnetic moment.[12,20] For the emergence of these complex magnetic properties, the coexistence of ferromagnetic and antiferromagnetic interactions in Cr-based magnets and the presence of the Cr-deficient layers are likely to be important. There has only been one previous study on the strains in magnetic fields of both compounds, which reported a large linear strain reaching 400 ppm under a magnetic field of 8.5 T in $Cr_2Te_3$.[7] Here, we report that $Cr_3Te_4$ and $Cr_2Te_3$ sintered samples show large magnetic-field-induced strains accompanied by large volume changes in a wide temperature region including room temperature. In particular, $Cr_3Te_4$ exhibits a large volume increase exceeding 1000 ppm by applying a magnetic field of 9 T at 300 K, which is larger than the reported values for other Cr-based magnets and comparable to that for invar alloys. Discussions based on the experimental data of the volume changes in magnetic fields, the temperature dependence of magnetization, and the thermal expansion suggest that the large magnetic-field-induced volume changes in both compounds are caused by a unique mechanism in which anisotropic lattice deformation and microstructural effects might play a role, which is different from the magneto-volume effect present in invar alloys.

Sintered samples of $Cr_3Te_4$ and $Cr_2Te_3$ were synthesized by a solid-state reaction method. Stoichiometric amounts of Cr and Te powders were mixed, pressed into pellets, and then sealed in an evacuated quartz tube. The tube was heated to and kept at 873 K for 12 h and then 1273 K for 48 h. The obtained polycrystalline samples were pulverized and sintered at 1073 K for 5 min using a spark plasma sintering furnace (SPS Syntex). Sample characterization was performed by powder X-ray diffraction (XRD) analysis with Cu K$\alpha$ radiation at room temperature using a MiniFlex diffractometer (RIGAKU), confirming that the single-phase samples of $Cr_3Te_4$ and $Cr_2Te_3$, with the same crystal structures as in previous studies, were obtained (Supplementary Fig.

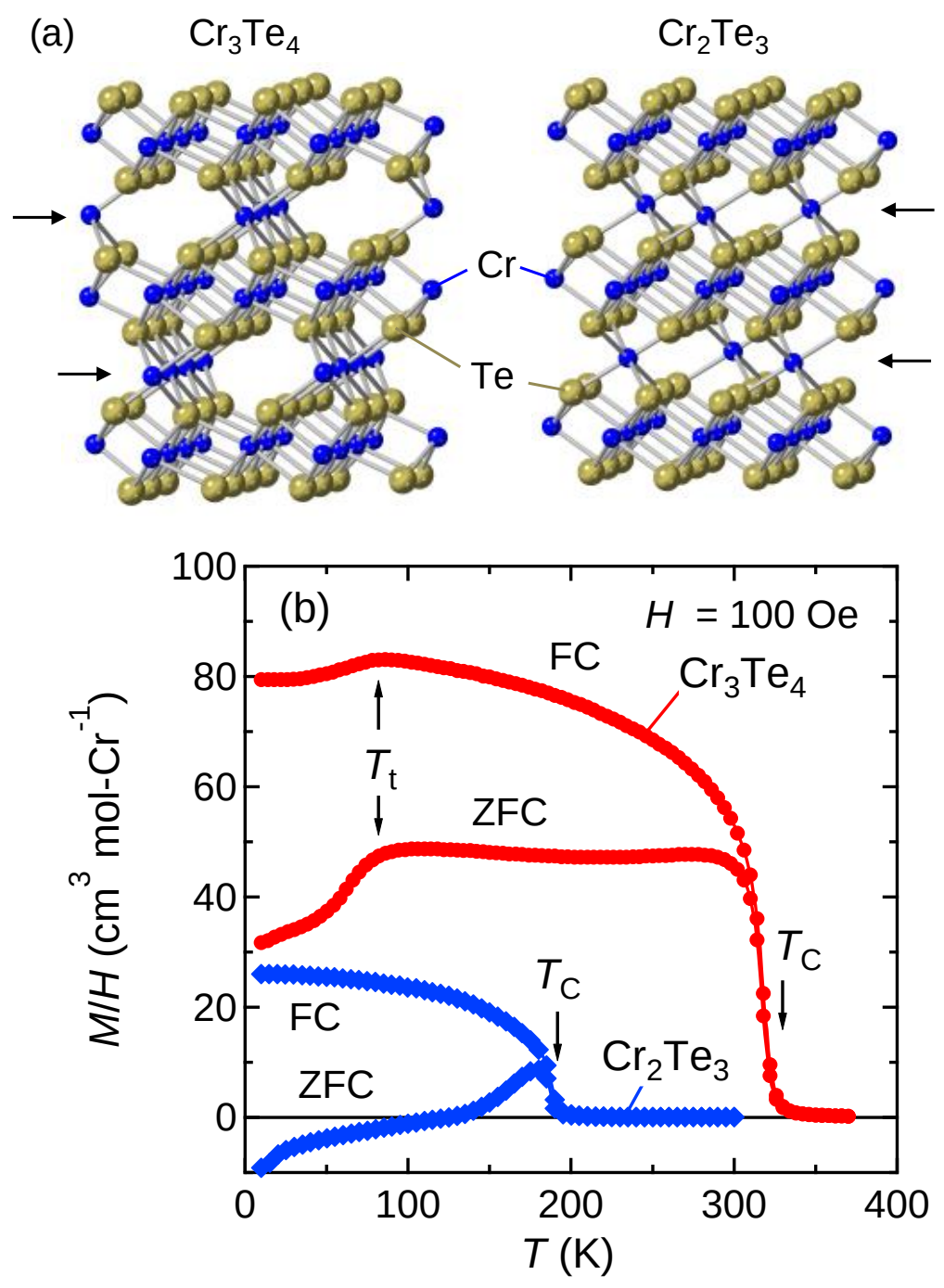

Fig. 1. (a) Crystal structures of $Cr_3Te_4$ (left) and $Cr_2Te_3$ (right). Arrows indicate the position of Cr-deficient layers. (b) Temperature dependence of zero-field-cooled (ZFC) and field-cooled (FC) magnetization of $Cr_3Te_4$ and $Cr_2Te_3$ sintered samples measured at 100 Oe.

1).[10,12] The linear strain in magnetic fields and linear thermal expansion of the sintered samples were measured using a strain gage (KFLB, 120W, Kyowa Electronic Instruments Co.) with a Cu reference.[21] The temperature control and application of the magnetic fields were achieved using a Physical Property Measurement System (Quantum Design). A part of the linear thermal expansion data was measured using a laser-interference dilatometer (LIX-2, Ulvac). In the isotropic situation, such as in the sintered sample, a volume change $\Delta V/V$ in a magnetic field, $H$, is represented as $\Delta V/V = 2(\Delta L/L)_\perp + (\Delta L/L)_{//}$, where $(\Delta L/L)_\perp$ and $(\Delta L/L)_{//}$ are the linear strains measured perpendicular and parallel to $H$, respectively. In the linear thermal expansion measurements of the isotropic samples, $\Delta V/V$ and $\Delta L/L$ have a relationship of $\Delta V/V = 3(\Delta L/L)$. Magnetization was measured using a Magnetic Property Measurement System (Quantum Design). As shown in Fig. 1(b), the synthesized $Cr_3Te_4$ and $Cr_2Te_3$ samples show ferromagnetic transitions at $T_C = 330$ and 190 K, respectively. The former also shows a slight decrease of magnetization at $T_t = 80$ K. Powder XRD patterns of $Cr_3Te_4$ and $Cr_2Te_3$ at various temperatures were measured using a SmartLab diffractometer (RIGAKU) with Cu K$\alpha_1$ radiation, which was monochromated by a Ge(111)-Johansson-type monochromator.

Figures 2(a) and (c) show the linear strains, $(\Delta L/L)_\perp$ and $(\Delta L/L)_{//}$, of the sintered samples of $Cr_3Te_4$ and $Cr_2Te_3$, respectively. First, the $(\Delta L/L)_\perp$ of $Cr_3Te_4$, shown in the left panel of Fig. 2(a), is always positive below 400 K and shows concave downward $H$ dependence below 350 K. The largest $(\Delta L/L)_\perp$ of 420 ppm was observed at $\mu_0 H = 9$ T and 50 K. In contrast, $(\Delta L/L)_{//}$ is positive between 0 and 9 T at 350 and 400 K, but below 300 K, $(\Delta L/L)_{//}$ takes negative values at low magnetic fields.

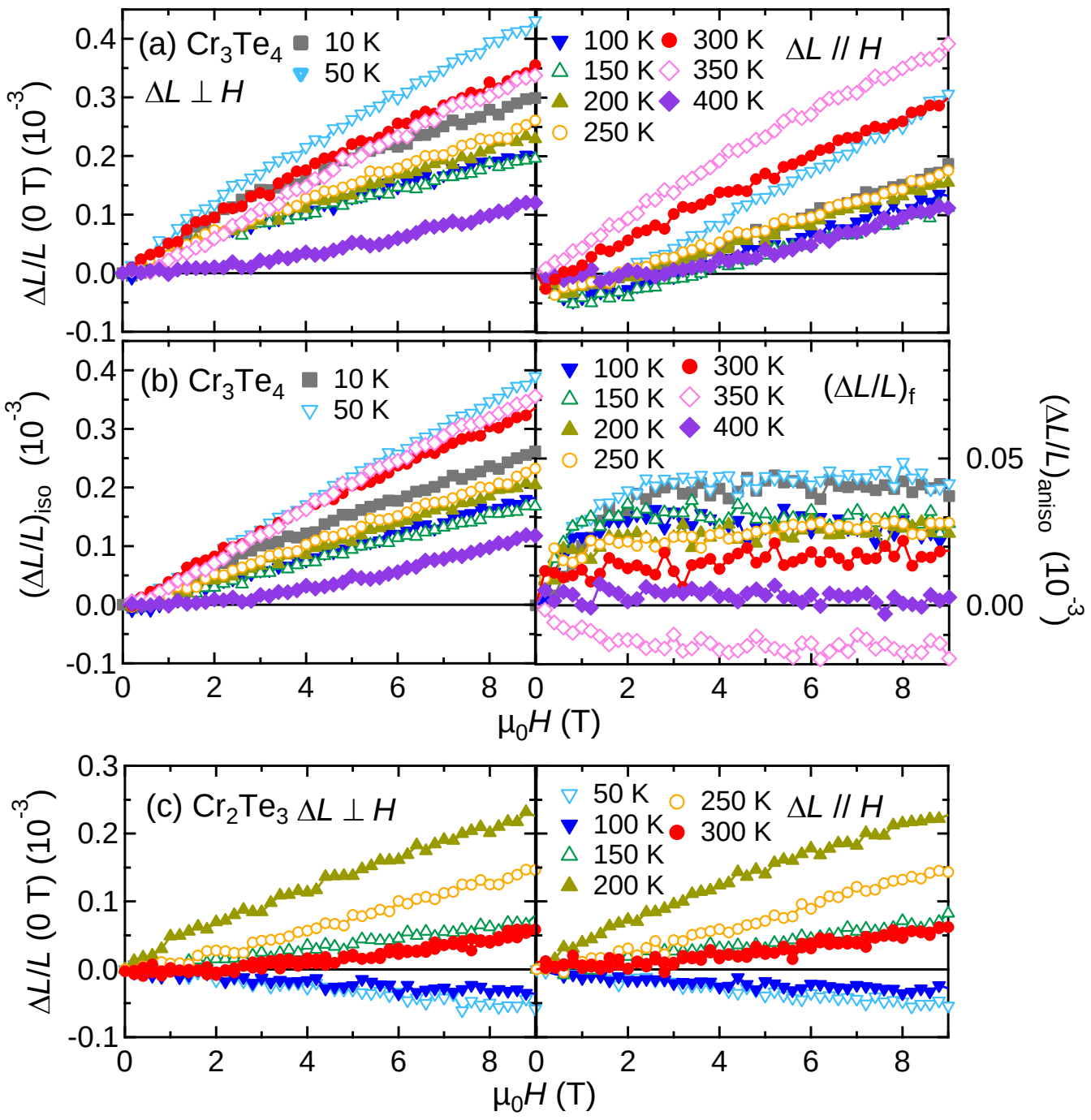

Fig. 2. Linear strains of (a) $Cr_3Te_4$ and (c) $Cr_2Te_3$ sintered samples measured in magnetic fields. All data were normalized to the zero-field data at each temperature. In (a) and (c), the left and right panels show the transverse and longitudinal magnetostrictions, $(\Delta L/L)_\perp$ and $(\Delta L/L)_{//}$, where linear strains were measured perpendicular and parallel to the magnetic field, respectively. (b) Isotropic and anisotropic contributions in the linear strains of $Cr_3Te_4$, defined as $(\Delta L/L)_{iso} = [2(\Delta L/L)_\perp + (\Delta L/L)_{//}]/3$ and $(\Delta L/L)_{aniso} = (\Delta L/L)_\perp - (\Delta L/L)_{iso}$, in the left and right panels, respectively.

This behavior of $(\Delta L/L)_\perp$ and $(\Delta L/L)_{//}$ can be explained using two contributions, namely, isotropic strain $(\Delta L/L)_{iso}$ and anisotropic strain $(\Delta L/L)_{aniso}$, as shown in Fig. 2(b). The former was defined as $(\Delta L/L)_{iso} = [2(\Delta L/L)_\perp + (\Delta L/L)_{//}]/3$, corresponding to the volume change of the sample. The latter was defined as $(\Delta L/L)_{aniso} = (\Delta L/L)_\perp - (\Delta L/L)_{iso} = -[(\Delta L/L)_{//} - (\Delta L/L)_{iso}]/2$, meaning the $\Delta L/L$ after the subtraction of the isotropic component $(\Delta L/L)_{iso}$. $(\Delta L/L)_{aniso}$ corresponds to the directional magnetostriction caused by the alignment of ferromagnetic domains in ferromagnets. Below 350 K, $(\Delta L/L)_{iso}$ almost linearly increased, whereas $(\Delta L/L)_{aniso}$ saturated below 2 T, corresponding to the magnetization curves (Supplementary Fig. 2). These results indicate that the analysis described above is appropriate and that the observed magnetic-field-induced strains consist of these two contributions. As shown in Fig. 3(a), $\Delta V/V = 3(\Delta L/L)_{iso}$ at 9 T reached a maximum value of 1170 ppm, as discussed below.

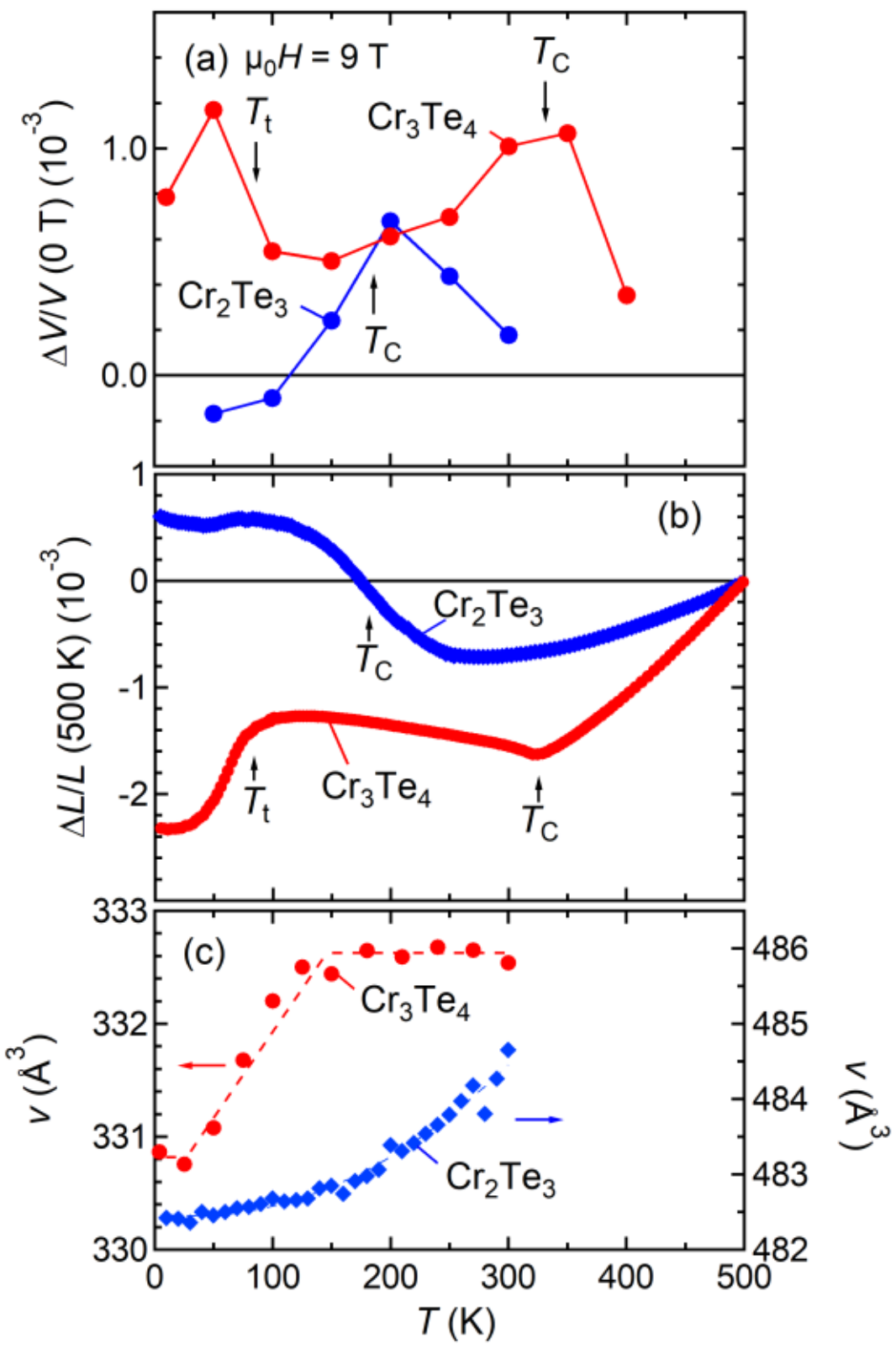

Fig. 3. (a) The volume changes of $Cr_3Te_4$ and $Cr_2Te_3$ sintered samples by applying a magnetic field of 9 T, estimated using the linear strain data shown in Fig. 2. (b) Linear thermal expansion of $Cr_3Te_4$ and $Cr_2Te_3$ sintered samples normalized to 500 K. (c) Temperature dependence of unit cell volumes of $Cr_3Te_4$ and $Cr_2Te_3$ sintered samples estimated using powder XRD data. In (c), the left and right axes show the data for $Cr_3Te_4$ and $Cr_2Te_3$, respectively.

In contrast, $(\Delta L/L)_{aniso}$ shows small values of several tens of ppm or less, which are normal values for a ferromagnet.

The linear strains of $Cr_2Te_3$ under the magnetic fields, shown in Fig. 2(c), are almost isotropic, i.e., $(\Delta L/L)_\perp \sim (\Delta L/L)_{//}$, unlike those of $Cr_3Te_4$. This means that $(\Delta L/L)_{aniso}$ in $Cr_2Te_3$ is negligibly small and most $\Delta L/L$ values consist of $(\Delta L/L)_{iso}$, accompanying the volume change. As shown in Fig. 3(a), the $\Delta V/V$ under the magnetic field of 9 T was a large value of 680 ppm at 200 K.

The observed magnetic-field-induced volume changes in the sintered samples of both Cr tellurides are reasonably large. In particular, the volume increases of $\Delta V/V = 1170$ ppm at 50 K and $\Delta V/V = 1070$ ppm at 350 K in $Cr_3Te_4$ under the magnetic field of 9 T are much larger than the maximum values of 780 and 730 ppm in $LiInCr_4S_8$ (22 K) and $AgCrS_2$ (42 K), respectively.[4,5] The maximum value of $\Delta V/V = 680$ ppm for $Cr_2Te_3$ under 9 T at 200 K is also comparable to these values. More importantly, all of the large volume changes in Cr-based magnets reported thus far appeared at low temperatures. However, $Cr_3Te_4$ shows large volume changes over a wide temperature region including room temperature. It is notable that a magnetic field of 9 T induced a significant volume change over 1000 ppm at room temperature and that the volume changes over 500 ppm occurred throughout the entire temperature range below 350 K.

We now discuss the mechanism of the large magnetic-field-induced strains accompanied by a large volume change in $Cr_3Te_4$ and $Cr_2Te_3$. As can be seen in Fig. 3(a), both materials show a particularly large $\Delta V/V$ near $T_C$, suggesting that the formation of long-range magnetic order is involved in the magnetic-field-induced volume change, same as in $LiInCr_4S_8$ and $AgCrS_2$.[4,5] However, in these Cr sulfides, large $\Delta V/V$ values only appeared in a very narrow temperature range of several K near $T_N$, whereas $Cr_3Te_4$ and $Cr_2Te_3$ showed large volume changes over a wide temperature range. In $LiInCr_4S_8$ and $AgCrS_2$, there is a discontinuous volume change at $T_N$ and the large magnetic-field-induced volume change is a consequence of a magnetic field effect on the phase transition at $T_N$. As shown in Fig. 3(b), the thermal expansion data of $Cr_3Te_4$ and $Cr_2Te_3$ also show large volume changes related to the magnetic ordering at $T_C$. However, the volume changes in Cr tellurides were gradual, unlike these sulfides, and appeared as negative thermal expansion. $Cr_3Te_4$ shows $\Delta V/V = -1080$ ppm from 120 to 330 K and $Cr_2Te_3$ shows $\Delta V/V = -3900$ ppm from 80 to 270 K with increasing temperature. In these temperature regions, the ferromagnetic order is not fully developed due to thermal fluctuations. Therefore, the application of magnetic fields facilitates the ferromagnetic order accompanied by the large volume increase, giving rise to a large volume increase induced by magnetic fields.

An important feature of the negative thermal expansion in Cr tellurides is the fact that the volume change in the dilatometric data does not match the crystallographic volume change determined by the diffraction experiments. As shown in Supplementary Fig. 3, the lattice constant $a$ of $Cr_3Te_4$ simply decreases with decreasing temperature from 300 K, whereas $b$ and $c$ increase above 125 K. As a result, the unit cell volume $v$ of $Cr_3Te_4$ shown in Fig. 3(c) is almost constant above 125 K and $dv/dT > 0$ below 125 K. In the case of $Cr_2Te_3$, $c$ decreases and $a$ increases with decreasing temperature, resulting in the gradual decrease of $v$. Thus, neither material has a temperature region with $dv/dT < 0$, consistent with previous studies.[22,23] Therefore, the negative thermal expansions in the dilatometric data shown in Fig. 3(b) are most likely due to microstructural effects, appearing in the sintered samples that exhibit anisotropic temperature dependence regarding their lattice constants.[24-26] The fact that the large magnetic-field-induced volume change and negative thermal expansion have the same origin of the volume change at the ferromagnetic order, strongly suggests that microstructural effects play an important role in the emergence of the large magnetic-field-induced volume change.

In contrast to the negative thermal expansion, to our knowledge, the magnetic-field-induced volume change due to microstructural effects has not been reported thus far. For negative thermal expansion due to microstructural effects, the presence of a direction along which the length of the unit cell shrinks with increasing temperature is essential ($b$ and $c$ axes for $Cr_3Te_4$ and $a$ axis for $Cr_2Te_3$).[25,26] Even when the crystallographic unit-cell volume exhibits positive thermal expansion, strong microstructural effect could realize negative thermal expansion in sintered samples.[27,28] Therefore, the emergence of negative thermal expansion in the temperature region with $dv/dT \geq 0$ in $Cr_3Te_4$ and $Cr_2Te_3$ suggests that strong microstructural effects are exhibited in these materials, implying that they could also be involved in the large magnetic-field-induced volume change. This is a different mechanism from the strong magneto-volume effect for invar alloys. The volume change by the microstructural effect can be strong against the repeated operation cycles and can also be effective in the fine particles,[29] suggesting that Cr tellurides are expected to be available as a magnetostrictive material.

Finally, we discuss the large $\Delta V/V$ that appears at 50 K in the $Cr_3Te_4$ data, which is probably related to the magnetic structure change at $T_t$ = 80 K.[12,18,19] As shown in Fig. 3(b), $Cr_3Te_4$ shows large positive thermal expansion with the volume change of $\Delta V/V$ = 3000 ppm at 50–80 K, indicating that the volume of the ferromagnetic phase at high temperatures is considerably larger than the canted antiferromagnetic phase at low temperatures. As in the case of the ferromagnetic transition at $T_C$, discussed above, the ferromagnetic state might become more stable than the canted antiferromagnetic state by applying a magnetic field, probably resulting in the considerable volume increase. However, the positive thermal expansion at 50–80 K also appear in the unit-cell volume, as shown in Fig. 3(c), suggesting that microstructural effects may not be involved in the large magnetic-field-induced volume change at 50–80 K. The large magnetic-field-induced volume change associated with the change of magnetic structure is similar to that in $LiFeCr_4O_8$, which exhibits a spin transition from ferrimagnetic to conical,[30] but the observed $\Delta V/V$ in $Cr_3Te_4$ at 50 K is more than double the maximum value in $LiFeCr_4O_8$.

In summary, the $Cr_3Te_4$ and $Cr_2Te_3$ sintered samples were found to show large strains accompanied by large volume increases in magnetic fields. The observed maximum volume increase under a magnetic field of 9 T is $\Delta V/V$ = 1170 and 680 ppm for $Cr_3Te_4$ and $Cr_2Te_3$, respectively. In particular, $Cr_3Te_4$ exhibited $\Delta V/V > 1000$ ppm at room temperature, which is comparable to the forced volume magnetostriction in invar alloys. Both the $Cr_3Te_4$ and $Cr_2Te_3$ samples showed a particularly large magnetic-field-induced volume change at around $T_C$ and negative thermal expansion caused by microstructural effects. The cooperation of anisotropic lattice deformation at the magnetic order and microstructural effects, which is in contrast to the magneto-volume effect operative in invar alloys, is most likely responsible for the large magnetic-field-induced volume changes in the Cr tellurides. This is a novel magneto-volume phenomenon occurring in Cr-based magnets.

## ACKNOWLEDGMENTS

This work was partly supported by JSPS KAKENHI (Grant Nos. 19H05823, 20H00346, and 20H02603) and the Iketani Science and Technology Foundation.

## Supplementary Note 1. Powder X-ray diffraction patterns of $Cr_3Te_4$ and $Cr_2Te_3$

Supplementary Fig. 1 shows the powder X-ray diffraction (XRD) patterns of the sintered samples of $Cr_3Te_4$ and $Cr_2Te_3$ measured at room temperature. All diffraction peaks in the $Cr_3Te_4$ data were indexed on the basis of monoclinic *C*2/*m* symmetry, while those in the $Cr_2Te_3$ data were indexed on the basis of trigonal $P\bar{3}1c$ symmetry, indicating that the single-phase samples of $Cr_3Te_4$ and $Cr_2Te_3$ with the same crystal structures as in previous studies were obtained.[10,12]

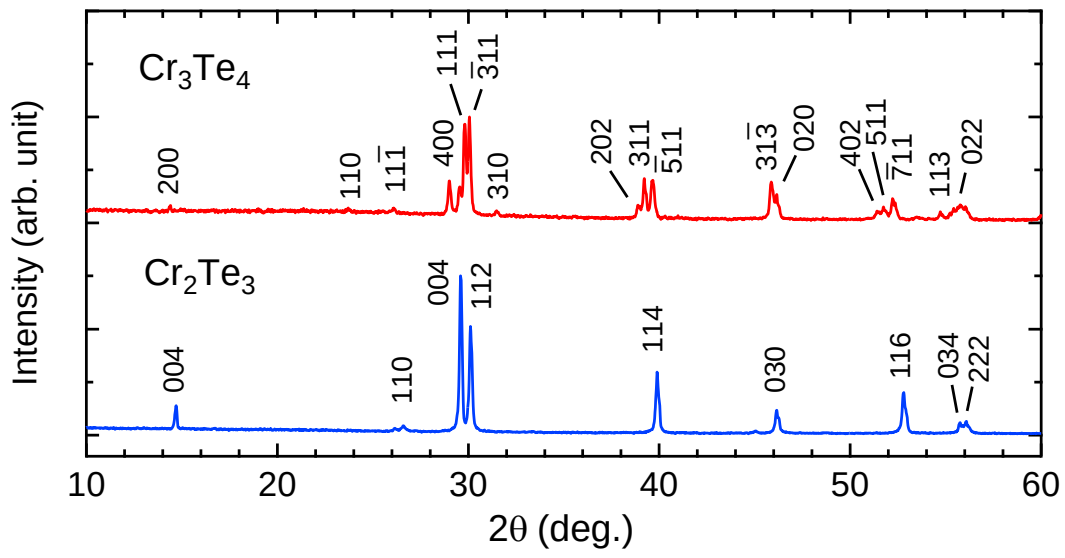

Supplementary Fig. 1. Powder XRD patterns of sintered samples of $Cr_3Te_4$ and $Cr_2Te_3$ measured at room temperature. Peak indices for the $Cr_3Te_4$ data are given using a monoclinic unit cell with lattice constants of $a$ = 13.9846 Å, $b$ = 3.9336 Å, $c$ = 6.8658 Å, and b = 118.302°, and those for $Cr_2Te_3$ are given using trigonal one with those of $a$ = 6.7986 Å and $c$ = 12.1076 Å.

## Supplementary Note 2. Magnetization process of $Cr_3Te_4$ sintered sample

Supplementary Fig. 2 shows the magnetization process of the $Cr_3Te_4$ sintered sample measured at 5, 50, 150, and 300 K. Magnetization measurements were performed using a Magnetic Property Measurement System (Quantum Design). All the data showed concave downward behavior, reflecting the presence of a ferromagnetic moment below $T_C$ = 330 K. The difference of the saturation field between 150 and 50 K might be due to the magnetic structure change at $T_t$.

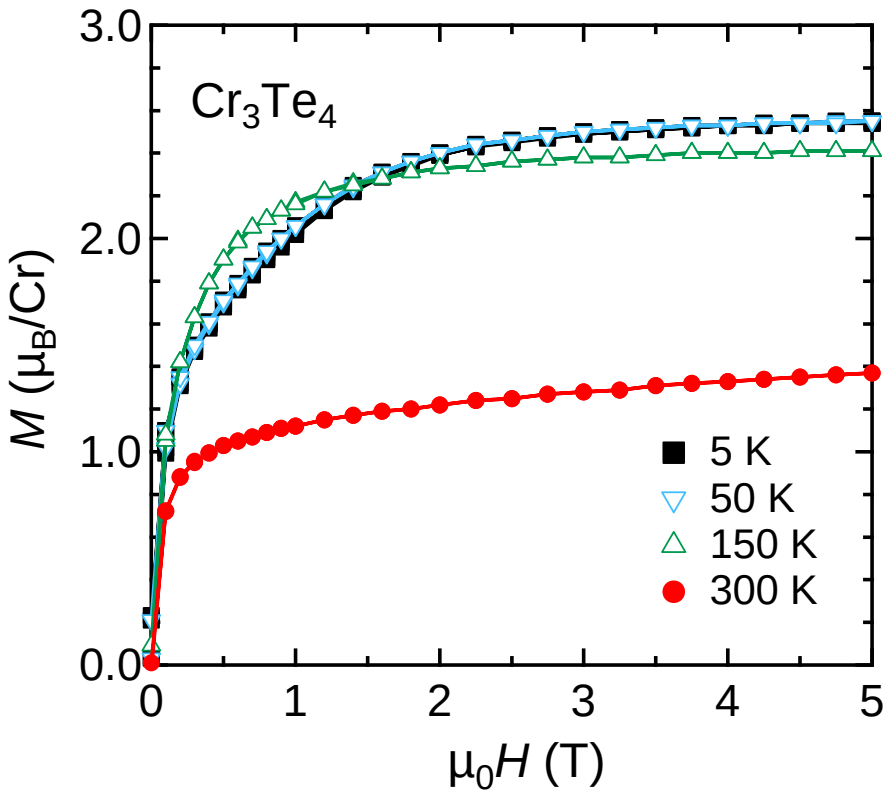

Supplementary Fig. 2. Magnetization curves of $Cr_3Te_4$ sintered samples measured at 5, 50, 150, and 300 K.

## Supplementary Note 3. Temperature dependences of lattice constants of $Cr_3Te_4$ and $Cr_2Te_3$.

Supplementary Fig. 3 shows the temperature dependences of the lattice constants determined by XRD data of the sintered samples of $Cr_3Te_4$ (a, b) and $Cr_2Te_3$ (c, d). In $Cr_3Te_4$, with decreasing temperature from 300 to 125 K, $a$ decreases while $b$ and $c$ increase, resulting in the almost constant unit cell volume $v$ above 125 K, as shown in Fig. 3(c). In $Cr_2Te_3$, $c$ decreases while $a$ increases with decreasing temperature in all the measured temperature range below 300 K, resulting in the gradual decrease of $v$, as shown in Fig. 3(c).

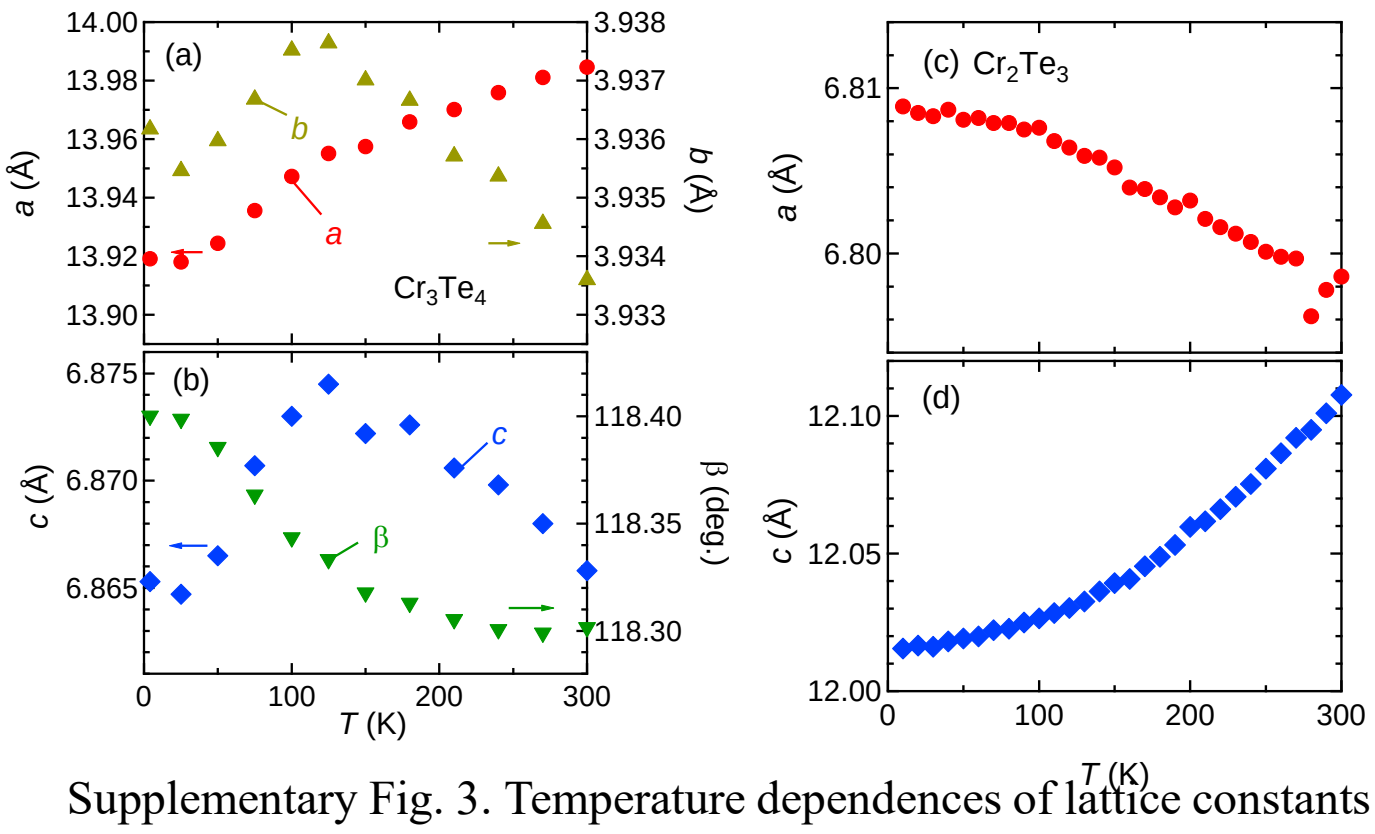

Supplementary Fig. 3. Temperature dependences of lattice constants determined by powder XRD measurements on $Cr_3Te_4$ (a, b) and $Cr_2Te_3$ (c, d) sintered samples. In (a) and (b), the left axis shows $a$ and $c$, while the right axis shows $b$ and b, respectively.